\documentclass[12pt]{article}
\usepackage{graphicx}

\begin{document}

\begin{center}
{\Large{\bf RADIATIVE DECAY OF \boldmath $\rho^0$ \unboldmath AND\\

\vspace{0.3cm}

\boldmath $\phi$ \unboldmath MESONS IN A CHIRAL UNITARY 

\vspace{0.3cm}

APPROACH
}}
\end{center}

\vspace{1cm}

\begin{center}
{\large{E.~Marco$^{1,2}$, S. Hirenzaki$^{3}$, E.~Oset$^{1,2}$
 and H. Toki$^{1}$}}
\end{center}

\begin{center}
{\small{$^{1}$ \it Research Center for Nuclear Physics,
Osaka University, Ibaraki,\\
Osaka 567-0047, Japan}}

{\small{$^{2}$ \it Departamento de F\'{\i}sica Te\'orica and IFIC, \\
Centro Mixto Universidad de Valencia-CSIC, \\
46100 Burjassot (Valencia), Spain}}

{\small{$^{3}$ \it Department of Physics, Nara Women's University,\\
 Nara 630-8506, Japan}}
\end{center}

\vspace{1cm}

\begin{abstract}
{\small{We study the $\rho^0$ and $\phi$ decays into $\pi^+ \pi^- \gamma$,
$\pi^0 \pi^0 \gamma$ and $\phi$ into $\pi^0 \eta \gamma$
using a chiral unitary approach to deal with
the final state interaction of the $M M$ system. The final state
interaction modifies only moderately the large momenta tail of the photon
spectrum of the $\rho^0 \rightarrow \pi^+ \pi^- \gamma$ decay. In
the case of $\phi$ decay the contribution to $\pi^+ \pi^- \gamma$
and $\pi^0 \pi^0 \gamma$ decay proceeds via kaonic loops
and gives a distribution of $\pi \pi$ invariant masses in
which the $f_0 (980)$ resonance shows up with a very distinct peak.
The spectrum found for $\phi\rightarrow\pi^0 \pi^0\gamma$ decay agrees with
the recent experimental results obtained at Novosibirsk.
The branching ratio for $\phi \rightarrow \pi^0 \eta \gamma$,
dominated by the $a_0 (980)$, is also in agreement with recent
Novosibirsk results.
}}
\end{abstract}

\vspace{2cm}


\newpage

In this work we investigate the reactions
$\rho \rightarrow \pi^+ \pi^- \gamma$, $\pi^0 \pi^0 \gamma$ and
$\phi \rightarrow \pi^+ \pi^- \gamma$, $\pi^0 \pi^0 \gamma$,
$\pi^0 \eta \gamma$,
treating the final state interaction of the two mesons with techniques of
chiral unitary theory recently developed. The energies of the two meson
system are too big in both the $\rho$ and $\phi$ decay to be treated with
standard chiral perturbation theory, $\chi PT$ \cite{GasLeu}. However,
a unitary coupled channels method, which makes use of the standard
chiral Lagrangians together with an expansion of Re $T^{-1}$ instead of the
$T$ matrix, has proved to be very efficient in describing the meson meson
interactions in all channels up to energies around 1.2 GeV
\cite{OllOse1,OllOsePel,GueOll}.
The method is analogous to the effective range expansion in
Quantum Mechanics. The work of \cite{GueOll} establishes a direct connection
with $\chi PT$ at low energies and gives the same numerical results
as the work of \cite{OllOsePel} where tadpoles and loops in the crossed
channels are not evaluated but are reabsorbed into the $L_i$ coefficients
of the second order Lagrangian of $\chi PT$. A technically much
simpler approach is done in \cite{OllOse1} where, only for $L=0$, it is shown
that the effect of the second order Lagrangian can be suitably
incorporated by means of the Bethe-Salpeter equation using the
lowest order Lagrangian as a source of the potential and a suitable
cut off, of the order of 1 GeV, to regularize the loops. This latter
approach will be the one used here, where the two pions
interact in s-wave.

The diagrammatic description for the $\rho \rightarrow \pi^+ \pi^-
\gamma$ decay is shown in Fig.~1

\begin{figure}[t]
\centerline{
\includegraphics[width=0.8\textwidth,angle=-90]{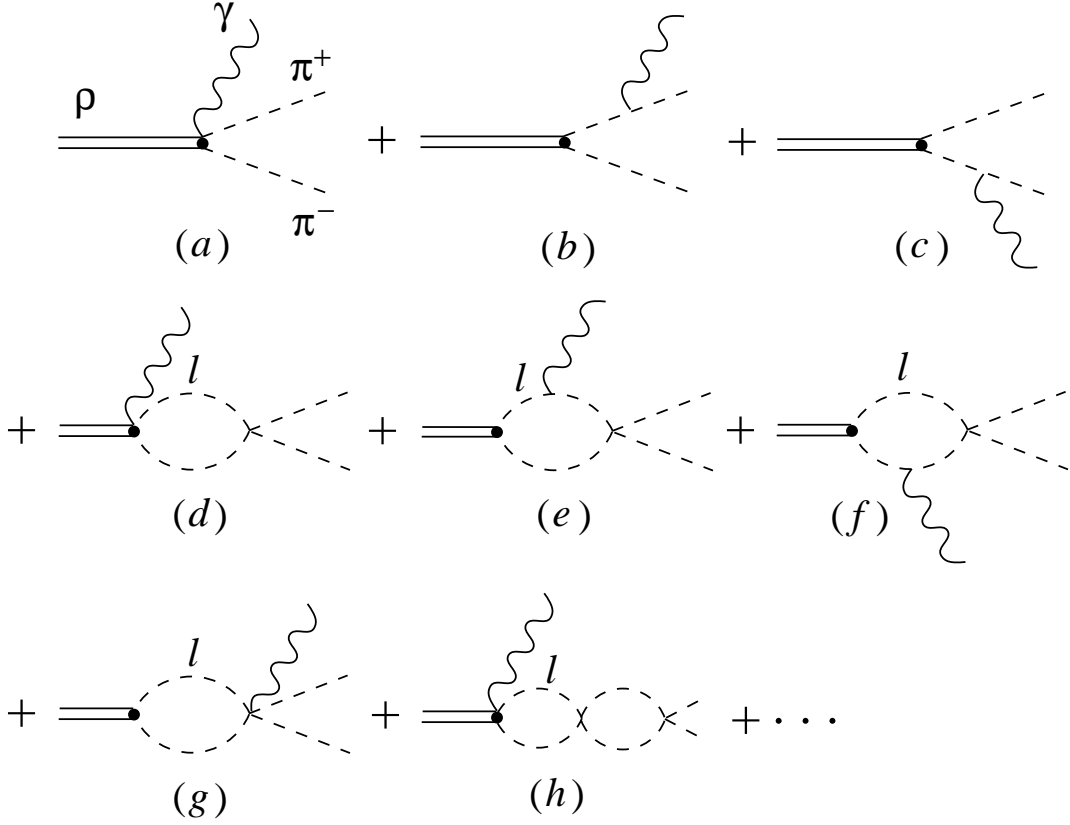}
}
\caption{Diagrams for the decay $\rho \rightarrow \pi^+ \pi^- \gamma$.
}
\label{fig1}
\end{figure}

In Fig.~1 the intermediate states in the loops attached to the
photon, $l$, can be $K^+ K^-$ or $\pi^+ \pi^-$. However, the
other loops involving only the meson meson interaction can be also
$K^0 \bar{K}^0$ or $\pi^0 \pi^0$ in the coupled channel approach of
\cite{OllOse1}.

For the case of $\pi^0 \pi^0 \gamma$ decay only the diagrams with
at least one loop contribute, $(d), (e), (f), (g), (h),\ldots$ in Fig.~1.

The case of the $\phi$ decay is analogous to the
$\rho \rightarrow \pi^0 \pi^0 \gamma$ decay. Indeed, the terms 
$(a), (b), (c)$ of Fig.~1 do not contribute since we do not have direct
$\phi \rightarrow \pi \pi$ coupling. Furthermore, there is another
novelty since only $K^+ K^-$ contributes to the loop with a
photon attached.

The procedure followed here in the cases of $\pi^0 \pi^0$ and $\pi^0 \eta$
production is
analogous to the one used in \cite{Bra}. Depending on the
renormalization scheme chosen, other diagrams can appear \cite{Bra}
but the whole set is calculated using gauge invariant arguments, as
done here, with the same result. The novelty in the present work is
that the strong interaction $ M M\rightarrow M'M'$ is evaluated using
the unitary chiral amplitudes instead of the lowest order used in \cite{Bra}.

We shall make use of the chiral Lagrangians for vector mesons of \cite{Ecker}
and follow the lines of ref.\ \cite{Huber} in the treatment of the radiative
rho decay. The Lagrangian coupling vector mesons to pseudoscalar mesons
and photons is given by

\begin{equation}             \label{eq:L2}
{\cal L}_2 [V(1^{- -})] = \frac{F_V}{2\sqrt{2}} \langle V_{\mu \nu}
f^{\mu \nu}_+ \rangle + \frac{i G_V}{\sqrt{2}} \langle V_{\mu \nu} u^{\mu}
u^{\nu} \rangle
\end{equation}
where $V_{\mu \nu}$ is a $3 \times 3$ matrix of antisymmetric tensor fields
representing the octet of vector mesons, $K^*$, $\rho$, $\omega_8$.
All magnitudes involved in Eq.\ (\ref{eq:L2}) are defined in \cite{Ecker}.
The coupling $G_V$ is deduced from the $\rho \rightarrow \pi^+ \pi^-$
decay and the $F_V$ coupling from $\rho \rightarrow e^+ e^-$.
We take the values chosen in \cite{Huber}, $G_V=67$ MeV, $F_V=153$ MeV.
The $\phi$ meson is introduced in the scheme by means of a singlet,
$\omega_1$, going from SU(3) to U(3) through the substitution
$V_{\mu \nu} \rightarrow V_{\mu \nu} +I_3 \frac{\omega_{1,\mu \nu}}{\sqrt{3}}$,
with $I_3$ the $3 \times 3$ diagonal matrix. Then, assuming ideal mixing
for the $\phi$ and $\omega$ mesons

\begin{eqnarray}
  \sqrt{\frac{2}{3}} \omega_1 + \frac{1}{\sqrt{3}} \omega_8 &\equiv& \omega
    \nonumber \\
  \frac{1}{\sqrt{3}} \omega_1 - \frac{2}{\sqrt{6}} \omega_8 &\equiv& \phi
\end{eqnarray}
one obtains the Lagrangian of Eq.\ (\ref{eq:L2}) substituting $V_{\mu \nu}$
by $\tilde V_{\mu \nu}$, given by

\begin{equation}
\tilde V_{\mu \nu} \equiv \left(\begin{array}{ccc} 
\frac{1}{\sqrt{2}} \rho^0_{\mu \nu} + \frac{1}{\sqrt{2}} \omega_{\mu \nu}
 & \rho^+_{\mu \nu} & K^{* +}_{\mu \nu} \\
\rho^-_{\mu \nu}& -\frac{1}{\sqrt{2}} \rho^0_{\mu \nu}
+ \frac{1}{\sqrt{2}} \omega_{\mu \nu} & K^{* 0}_{\mu \nu} \\
K^{* -}_{\mu \nu} & \bar{K}^{*0}_{\mu \nu} & \phi_{\mu \nu}
\end{array}
\right)
\end{equation}
From there one can obtain the couplings corresponding to $VPP$ ($V$ vector
and $P$ pseudoscalar) and $VPP\gamma$ with the $G_V$ term or the
$VPP\gamma$ with the $F_V$ term.

The basic couplings needed to evaluate the diagrams of Fig.~1 are

\begin{eqnarray}
t_{\rho \pi^+ \pi^-} &=& -\frac{G_V M_{\rho}}{f^2} (p_{\mu}-p'_{\mu})
\epsilon^{\mu} (\rho) \nonumber \\
t_{\rho \gamma \pi^+ \pi^-} &=& 
2 e \frac{G_V M_{\rho}}{f^2} \epsilon_{\nu} (\rho) \epsilon^{\nu} (\gamma)
\nonumber \\
&&+ \frac{2e}{M_{\rho} f^2} \left(\frac{F_V}{2} -G_V\right)
P_{\mu} \epsilon_{\nu} (\rho) 
[k^{\mu} \epsilon^{\nu} (\gamma) -k^{\nu} \epsilon^{\mu} (\gamma)]
\label{eq:couplings}\\
t_{\gamma \pi^+ \pi^-} &=& 2 e p_{\mu} \epsilon^{\mu} (\gamma) \nonumber
\end{eqnarray}
with $p_{\mu}$, $p'_{\mu}$ the $\pi^+, \pi^-$ momenta, $P_{\mu}$, 
$k_{\mu}$ the $\rho$ and photon momenta and $f$ the pion decay
constant which we take as $f_\pi=93$ MeV. 

The vertices of Eq.\ (\ref{eq:couplings})
are easily generalized to the case of $K^+ K^-$. Using the Lagrangian of
Eq.\ (\ref{eq:L2}), in the first two couplings one has an extra factor
$1/2$ and the last coupling is the same. The couplings for
$\phi K^+ K^-$ and $\phi \gamma K^+ K^-$ which are needed for the
$\phi$ decay are like the two first couplings of Eq.~(\ref{eq:couplings})
substituting $M_{\rho}$ by $M_{\phi}$, $\epsilon^{\mu} (\rho)$
by $\epsilon^{\mu} (\phi)$ and multiplying by $-1/\sqrt{2}$.
In addition we shall take the values $G_V=55$ MeV and $F_V=165$ MeV
which are suited to the $\phi \rightarrow K^+ K^-$ and 
$\phi \rightarrow e^+ e^-$ decay widths respectively.

The evaluation of the $\rho$ width for the first three diagrams
$(a), (b), (c)$ of Fig.~1 is straightforward and has been done before
\cite{BaiKho,Singer,Renard} and in \cite{Huber} following the present
formalism. We rewrite the results in a convenient way for our purposes

\begin{eqnarray}
\frac{d\Gamma_{\rho}}{d M_I} &=& \frac{1}{(2\pi)^3}
\frac{1}{16 m_{\rho}^3} (m_{\rho}^2 - M_I^2)(M_I^2 - 4 m_{\pi}^2)^{1/2}
\nonumber \\
&&\times \frac{1}{2} \int_{-1}^1 d\cos \theta \bar{\sum} \sum |t|^2
\label{eq:dgammadmi}
\end{eqnarray}
where

\begin{equation}        \label{eq:sumt2}
\bar{\sum} \sum |t|^2 = \frac{8}{3} e^2 \left[I_1 + I_2 +
I_3\right]
\end{equation}
In Eq.~(\ref{eq:dgammadmi}), $M_I$ is the invariant mass of the
two $\pi$ system and $\theta$ the angle between the $\pi^+$ meson
and the photon in the frame where the $\pi^+ \pi^-$ system
is at rest. The quantity $I_1$ stands
for the contribution of the first diagram alone, Fig.~1 $(a)$, $I_3$ for
the second and third $(b),(c)$ and $I_2$ for the interference between
the first diagram and the other two. They are given by

\begin{eqnarray}        \label{eq:I1I2I3}
I_1 & = & \left|\frac{M_{\rho} G_V}{f^2} + \frac{K}{f^2}
\left(\frac{F_V}{2}-G_V\right)\right|^2  \nonumber \\
I_2 & = & 2 \frac{M_{\rho} G_V}{f^2} \vec{p}\,^2 (D_1 + D_2)\sin^2 \theta 
\left\{ \frac{M_{\rho} G_V}{f^2} + \frac{K}{f^2}
\left(\frac{F_V}{2}-G_V\right)\right\}
 \nonumber\\
I_3 & = & 2 \vec{p}\,^2  (D_1 + D_2)\sin^2 \theta \nonumber \\
&& \times \left\{ (D_1 + D_2)
\vec{p}\,^2 + (D_1 - D_2) |\vec{p}|  |\vec{k}| \cos \theta 
\right\} \left(\frac{M_{\rho} G_V}{f^2} \right)^2
\end{eqnarray}
where $K$ is the photon momentum in the $\rho$ rest frame and 
$p, k$ are the momenta of the meson and the photon in the
rest frame of the $\pi^+ \pi^-$ system, and $D_1, D_2$ the meson
propagators in the $(b), (c)$ Bremsstrahlung diagrams, conveniently
written in terms of $M_I$ and $\theta$.

The first term of the contact term, $t_{\rho \gamma \pi^+ \pi^-}$,
in Eq.\ (\ref{eq:couplings}) is not gauge invariant. It requires
the addition of the diagrams $(b)$ and $(c)$ of
Fig.\ 1 to have a gauge invariant set. On the other hand the second
term in the contact term ($F_V/2-G_V$ part) is gauge
invariant by itself. When considering final state interaction of the
mesons this means that the $G_V$ part of the contact term, diagram $(d)$,
must be complemented by diagrams $(e)$, $(f)$, $(g)$ to form the gauge
invariant set. On the other hand the $F_V/2-G_V$ part of the
contact term appears in the $(d)$ diagram which is gauge invariant by
itself.

The technology to introduce the final state interactions is available
from the study of $\phi \rightarrow K^0 \bar{K}^0 \gamma$ in
\cite{Oller}. There it was shown that the strong $t$ matrix for the
$M_1 M_2 \rightarrow M'_1 M'_2$ transition factorizes with their
on shell values in the loops with a photon attached. The same
was proved for the loops of the Bethe-Salpeter equation in the
meson meson interaction description of \cite{OllOse1}. On the other hand
the sum of the diagrams $(d), (e), (f), (g)$, which appears now with the $G_V$
part of the contact term (diagram $(a)$), could be done using
arguments of gauge invariance which led to a finite contribution
for the sum of the loops \cite{Bra,LucPer,CloIsgKum}. A sketch of the
procedure is given here. The $\rho \rightarrow \pi^+ \pi^- \gamma$
amplitude can be written as $\epsilon_{\mu}(\rho) \epsilon_{\nu}(\gamma)
T^{\mu \nu}$ and the structure of the loops in Fig.\ 1 is such that

\begin{equation}            \label{eq:Tmunu}
T^{\mu \nu} = a \, g^{\mu \nu} + b\, Q^{\mu} Q^{\nu} + c\, Q^{\mu} K^{\nu}
+ d\, Q^{\nu} K^{\mu} + e\, K^{\mu} K^{\nu}
\end{equation}
where $Q$, $K$ are the $\rho$ meson and photon momenta respectively.
Gauge invariance $(T^{\mu \nu} K_{\nu}=0)$ forces $b = 0$ and
$d = -a/(Q\cdot K)$. Furthermore, in the Coulomb gauge only the
$g^{\mu \nu}$ term of Eq.(\ref{eq:Tmunu}) contributes and the
coefficient $a$ is calculated from the $d$ coefficient, to which
only the diagrams $(e)$, $(f)$, of Fig.\ 1 contribute. For dimensional
reasons the loop integral contains two powers less in the internal
variables than the pieces contributing to the $g^{\mu \nu}$ term
from these diagrams, since the product $Q^{\nu} K^{\mu}$ is factorized
out of the integral. This makes the $d$ coefficient finite. Furthermore,
the $M M \rightarrow M M$ vertices appearing there have the structure
$\alpha s + \beta \sum_i p_i^2 + \gamma \sum_i m_i^2$, which can be
recast as $\alpha s + (\beta + \gamma) \sum_i m_i^2 +
\beta \sum_i (p_i^2 - m_i^2)$. The first two terms in the sum give the
on shell contribution and the third one the off shell part. This latter
term kills one of the meson propagators in the loops and does not
contribute to the $d$ term in Eq.\ (\ref{eq:Tmunu}). Hence, the meson meson
amplitudes factorize outside the loop integral with their on shell
values. A more detailed description, done for a similar problem,
can be seen in \cite{photomprc}, following the steps from Eqs.\ (13)
to (23).

Following these steps, as done in \cite{Oller,photomprc},
it is easy to include the effect of the final state
interaction of the mesons. The sum of the diagrams
$(d), (e), (f), (g)$ and further iterated loops of the meson-meson
interaction, $(h), \ldots,$ is shown to have the same structure
as the contact term of $(a)$ in the Coulomb gauge, which
one chooses to evaluate the amplitudes. The sum of all terms
including loops is readily accomplished by multiplying the
$G_V$ part of the contact term by the factor $F_1(M_{\rho}, M_I)$

\begin{equation}        \label{eq:f1mrhomi}
F_1(M_{\rho}, M_I) = 1 + \tilde G_{\pi^+ \pi^-} 
t_{\pi^+ \pi^-, \pi^+ \pi^-} + 
\frac{1}{2} \tilde G_{K^+ K^-} t_{K^+ K^-, \pi^+ \pi^-} 
\end{equation}
where $t_{M_1 M_2,M'_1 M'_2}$ are the strong transition matrix
elements in s-wave evaluated in \cite{OllOse1} and $\tilde G_{M_1 M_2}$
is given by

\begin{eqnarray}
\tilde G_{M_1 M_2} (M_{\rho}, M_I) = \frac{1}{8 \pi^2} (a-b)
I(a,b) \nonumber \\
a= \frac{M_{\rho}^2}{M_{M_1}^2}; \quad b= \frac{M_I^2}{M_{M_1}^2}
\end{eqnarray}
with $I(a,b)$ a function given analytically in \cite{Oller}.
The  $({F_V}/{2}-G_V)$ part of the contact term is iterated by
means of diagrams $(d)$, $(h)\ldots$ in order to account for final state
interaction. Here the loop function is the ordinary two meson propagator
function, $G$, of the Bethe-Salpeter equation, $T= V+VGT$, for the meson-meson
scattering and which is regularized in \cite{OllOse1} by means of a
cut-off in order to fit the scattering data. The sum of all these
diagrams is readily accomplished by multiplying the $({F_V}/{2}-G_V)$
part of the contact term by the factor

\begin{equation}        \label{eq:f2mi}
F_2(M_I) = 1 + G_{\pi^+ \pi^-} 
t_{\pi^+ \pi^-, \pi^+ \pi^-} + 
\frac{1}{2} G_{K^+ K^-} t_{K^+ K^-, \pi^+ \pi^-} 
\end{equation}
By using isospin Clebsch Gordan coefficients the amplitudes
$t_{M_1 M_2,M'_1 M'_2}$ can be written in terms of the isospin
amplitudes of \cite{OllOse1} as

\begin{eqnarray}        \label{eq:tmmmm}
t_{\pi^+ \pi^-,\pi^+ \pi^-} & = & \frac{2}{3}
 t_{\pi \pi,\pi \pi}^{I=0} (M_I) \nonumber\\
t_{K^+ K^-,\pi^+ \pi^-}&=&\frac{1}{\sqrt{3}}t_{K \bar{K},\pi \pi}^{I=0} (M_I)
\end{eqnarray}
neglecting the small $I=2$ amplitudes. In Eq.~(\ref{eq:tmmmm}), one
factor $\sqrt{2}$ for each $\pi^+ \pi^-$ state has been introduced, since
the isospin amplitudes of \cite{OllOse1} used in Eq.~(\ref{eq:tmmmm})
are written in a unitary normalization which includes
an extra factor $1/\sqrt{2}$ for each $\pi \pi$ state.

The invariant mass distribution in the presence of final
state interaction is now given by
Eqs.~(\ref{eq:dgammadmi}, \ref{eq:sumt2}, \ref{eq:I1I2I3}) by
changing in Eq.~(\ref{eq:I1I2I3})

\begin{eqnarray}        \label{eq:I1I2I3FSI}
I_1 & \rightarrow & \left|\frac{M_{\rho} G_V}{f^2} F_1(M_{\rho},M_I)
+ \frac{K}{f^2} \left(\frac{F_V}{2}-G_V\right) F_2(M_I)\right|^2
 \nonumber \\
I_2 & \rightarrow & 2 \frac{M_{\rho} G_V}{f^2}
\vec{p}\,^2 (D_1 + D_2)\sin^2 \theta \nonumber\\
&&\times \mbox{Re}\left\{ \frac{M_{\rho} G_V}{f^2} F_1(M_{\rho},M_I)
+ \frac{K}{f^2} (\frac{F_V}{2}-G_V) F_2(M_I)\right\}
 \nonumber\\
I_3 & \rightarrow & I_3
\end{eqnarray}
The $\rho \rightarrow \pi^0 \pi^0 \gamma$ width is readily obtained
by omitting the terms $I_2, I_3$ and also omitting the first
term (the unity) in the definition of the $F_1(M_{\rho}, M_I)$,
$F_2(M_I)$ factors in Eqs.~(\ref{eq:f1mrhomi}) and (\ref{eq:f2mi})
and dividing by a factor two the width to account for the identity of the
particles.

The evaluation of the $\phi$ decay is straightforward by noting
that the tree level contributions, diagrams $(a), (b), (c)$
are not present now, and that only kaonic loops attached to photons
contribute in this case. Hence, the invariant mass distribution for
$\phi \rightarrow \pi^+ \pi^- \gamma$ is given in this case by
Eq.~(\ref{eq:dgammadmi}), changing $m_{\rho} \rightarrow m_{\phi}$,
with

\begin{equation}        
\bar{\sum} \sum |t|^2 = \frac{4}{3} e^2 
\left| \frac{M_{\phi} G_V}{f^2} \frac{1}{\sqrt{3}} \tilde G_{K^+ K^-}
t_{K \bar{K}, \pi \pi}^{I=0} 
+ \frac{K}{f^2} \left(\frac{F_V}{2} - G_V\right) \frac{1}{\sqrt{3}} G_{K^+ K^-}
t_{K \bar{K}, \pi \pi}^{I=0} \right|^2
\end{equation}
For $\phi \rightarrow \pi^0 \pi^0 \gamma$ the cross section is the same
divided by a factor two to account fot the identity of the two
$\pi^0$'s.

For the $\phi \rightarrow \pi^0 \eta \gamma$ case we have

\begin{equation}        
\bar{\sum} \sum |t|^2 = \frac{4}{3} e^2  
\left| \frac{M_{\phi} G_V}{f^2} \frac{1}{\sqrt{2}} \tilde G_{K^+ K^-}
t_{K \bar{K}, \pi \eta}^{I=1} 
+ \frac{K}{f^2} \left(\frac{F_V}{2} - G_V\right) \frac{1}{\sqrt{2}}
\tilde G_{K^+ K^-}
t_{K \bar{K}, \pi \eta}^{I=1}
\right|^2
\end{equation}

\begin{figure}[t]
\centerline{
\includegraphics[width=0.6\textwidth,angle=-90]{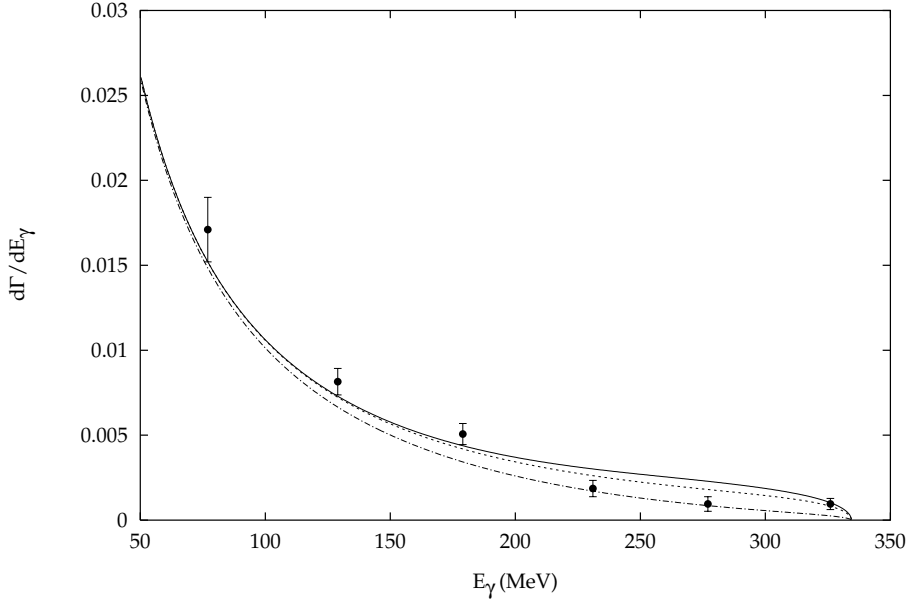}
}
\caption{Photon distribution, $d\Gamma/dK$, for the process
$\rho \rightarrow \pi^+ \pi^- \gamma$ as a function of
the photon momentum. Solid line: spectrum including
final state interaction of the two mesons and the $F_V$ and $G_V$
contributions; dashed line: spectrum including
only the tree level diagrams $(a), (b), (c)$ of Fig.~1 and
the $F_V$ and $G_V$ contributions; dashed-dotted line: spectrum including
only the tree level diagrams $(a), (b), (c)$ of Fig.~1 and
taking $F_V = 0$. The experimental
data taken from \cite{Dolinsky} are normalized to our results.
}
\label{fig2}
\end{figure}

In Fig.~2 we show $d\Gamma/dK$ for $\rho \rightarrow \pi^+ \pi^- \gamma$
decay, $(d\Gamma_{\rho}/dK = m_{\rho} d\Gamma_{\rho}/M_I dM_I)$.
The dashed-dotted line shows the contribution of diagrams
$1(a), (b), (c)$ and taking $F_V = 0$. The dashed line shows again
the contribution coming from diagrams $1(a), (b), (c)$ but now considering
also the $F_V$ contributions. Finally, the solid line includes the full set
of diagrams in Fig.\ 1 to account for final state interaction and
with the $F_V$ and $G_V$ contributions.
The process is infrared divergent and we plot the distribution
for photons with energy bigger than 50 MeV, where the experimental
measurements exist \cite{Dolinsky}. We have also added the experimental
data, given in \cite{Dolinsky} with arbitrary normalization, normalized
to our results.

As one can see in Fig.~2, the shape of the distribution of photon
momenta is well reproduced. For the total contribution
we obtain a branching ratio to the total
width of the $\rho$

\begin{equation}
B(\rho^0 \rightarrow \pi^+ \pi^- \gamma) = 1.18\,\, 10^{-2}\quad
\mbox{for $K>50$ MeV}
\end{equation}
which compares favourably with the experimental number
\cite{Dolinsky}, $B^{\mbox{\scriptsize exp}} 
(\rho^0 \rightarrow \pi^+ \pi^- \gamma)
= (0.99 \pm 0.04 \pm 0.15)\,\, 10^{-2}$ for $K>50$ MeV.

The changes induced by the $F_V$ term found here reconfirm
the findings of \cite{Huber}.
The effect of the final state interaction is small and mostly
visible at high photon energies, where it increases
$d\Gamma/dK$ by about 25\%. The branching ratio for
$B(\rho^0 \rightarrow \pi^0 \pi^0 \gamma)$ that we obtain
is $1.4 \,\,10^{-5}$ which can be interpreted in our case as
$\rho^0 \rightarrow \gamma \sigma(\pi^0 \pi^0)$ since the
$\pi^0 \pi^0$ interaction is dominated by the $\sigma$ pole
in the energy regime where it appears here. This result is very similar
to the one obtained in \cite{Bra}. In the case one
considers $F_V G_V<0$, the result obtained is
$1.0 \,\,10^{-4}$. The measurement of this quantity may serve as a test
for the sign of the $F_V G_V$ product.

\begin{figure}[t]
\centerline{
\includegraphics[width=0.6\textwidth,angle=-90]{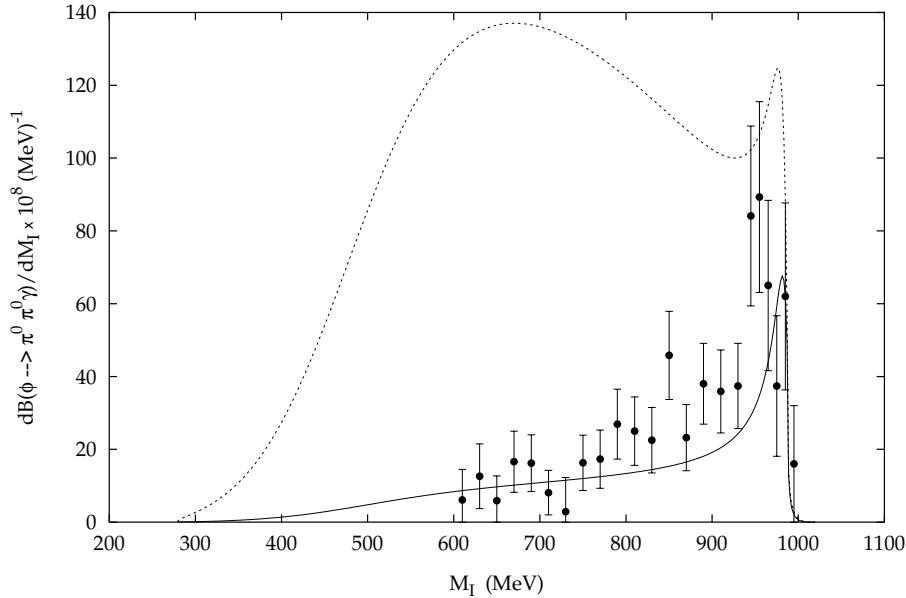}
}
\caption{Distribution $dB /dM_I$ for the decay
$\phi \rightarrow \pi^0 \pi^0 \gamma$, with $M_I$ the invariant mass of
the $\pi^0 \pi^0$ system. Solid line: our prediction, with $F_V G_V >0$.
Dashed line: result taking $F_V G_V <0$.
The data points are from \cite{Novo} and only
statistical errors are shown. The systematic errors are similar to the
statistical ones \cite{Novo}. The distribution for $\phi \rightarrow \pi^+
\pi^- \gamma$ is twice the results plotted there.
}
\label{fig3}
\end{figure}

As for the $\phi\rightarrow \pi \pi \gamma$ decay, as we pointed above
the $\phi\rightarrow \pi^+ \pi^- \gamma$ rate is twice the one of the
$\phi\rightarrow \pi^0 \pi^0 \gamma$. We have evaluated the invariant
mass distribution for these decay channels and in Fig.~3 we plot
the distribution $dB/dM_I$ for $\phi \rightarrow \pi^0 \pi^0 \gamma$ 
which allows us to see the
$\phi \rightarrow f_0 \gamma$ contribution since the $f_0$ is
the important scalar resonance appearing in the
$K^+ K^- \rightarrow \pi^+ \pi^-$ amplitude \cite{OllOse1}. 
The solid curve shows our prediction, with $F_V G_V >0$, the sign
predicted by vector meson dominance \cite{Ecker}.
The dashed curve is obtained considering  $F_V G_V <0$.
We compare
our results with the recent ones of the Novosibirsk experiment \cite{Novo}.
We can see that the shape of the spectrum is relatively well reproduced
considering statistical and systematic errors (the latter ones not shown in
the figure). The results considering $F_V G_V <0$ are
in complete disagreement with the data.

The finite total branching ratio which we find for 
$\phi \rightarrow \pi^+ \pi^- \gamma$ is $1.6 \,\,10^{-4}$ and correspondingly
$0.8 \,\,10^{-4}$ for the $\phi \rightarrow \pi^0 \pi^0 \gamma$. This latter
number is slightly smaller than
the result given in  \cite{Novo},
$(1.14\pm 0.10\pm 0.12)\,\,10^{-4}$, where the first error is statistical
and the second one systematic. The result given in
\cite{CMDpi0pi0} is
$(1.08\pm 0.17\pm 0.09)\,\,10^{-4}$, compatible with our prediction.
The branching ratio measured in \cite{CMDpi+pi-} for 
$\phi \rightarrow \pi^+ \pi^- \gamma$ is $(0.41\pm 0.12\pm 0.04)\,\,10^{-4}$.

The branching ratio obtained for the case
$\phi \rightarrow \pi^0 \eta \gamma$ is $0.87 \,\,10^{-4}$. The
results obtained at Novosibirsk are \cite{pi0eta} $(0.83 \pm 0.23)\,\,10^{-4}$
and \cite{CMDpi0pi0}
$(0.90\pm 0.24\pm 0.10)\,\,10^{-4}$. The spectrum, not shown,
is dominated by the $a_0$ contribution.

The contribution of $\phi \rightarrow f_0(\pi^+ \pi^-) \gamma$, obtained by
integrating $d\Gamma_{\phi} /dM_I$ assuming an approximate
Breit-Wigner form to the left of the $f_0$ peak, gives us a branching
ratio $0.44 \,\,10^{-4}$. As argued above, the branching ratio for
$\phi \rightarrow \pi^0 \pi^0 \gamma$ is one half of 
$\phi \rightarrow \pi^+ \pi^- \gamma$, which should not be compared to the one
given in  \cite{Novo} since there the assumption that all the strength of the
spectrum is due to the $f_0$ excitation is done. As one can see in Fig.~3, we 
find also an appreciable strength for $\phi\rightarrow\sigma\gamma$.

We should also warn not to compare our predicted rate for
$\phi \rightarrow \pi^+ \pi^- \gamma$ directly with experiment. Indeed,
the experiment is done using the reaction
$e^+ e^- \rightarrow \phi \rightarrow \pi^+ \pi^- \gamma$, which interferes
with the $\rho$ contribution 
$e^+ e^- \rightarrow \rho \rightarrow \pi^+ \pi^- \gamma$
at the tail of the $\rho$ mass distribution in the $\phi$ mass region
\cite{BraColGre}. Also the results in \cite{CMDpi0pi0,CMDpi+pi-} are based on
model dependent assumptions. For these reasons, as quoted in \cite{CMDpi0pi0},
the $\pi^0 \pi^0 \gamma$ mode is more efficient to study the
$\pi \pi$ mass spectrum.

Our result for $\phi \rightarrow \pi^0 \pi^0 \gamma$ 
is 50 \% larger than the one obtained in \cite{Bra} owed
to the use of the unitary $K^+ K^-
\rightarrow\pi^0 \pi^0$ amplitude instead of the lowest order chiral one. The
shape of the distribution found here is, however, rather different than the one
obtained in \cite{Bra}, showing the important contribution of the $f_0$
resonance which appears naturally in the unitary chiral approach. 

The $\phi \rightarrow f_0 \gamma$ decay has been advocated
as an important source of information on the nature of the
$f_0$ resonance and experiments have been conducted at
Novosibirsk \cite{Akhmetshin} and are also planned at Frascati
\cite{FraKimLee}, trying to magnify the signal for $f_0$ production through
interference with initial and final state radiation in the
$e^+ e^- \rightarrow \phi \rightarrow f_0(\pi^+ \pi^-) \gamma$ reaction
\cite{BraColGre,FraKimLee,ColFra,AchGabSol}. The completion of the experiments
\cite{Novo,CMDpi0pi0,pi0eta,CMDpi+pi-} is a significant step forward.

Present evaluations for
$\phi \rightarrow f_0 \gamma \rightarrow \pi \pi \gamma$
are based on models assuming a $K \bar{K}$ molecule for
the $f_0$ \cite{AchIva} with a branching ratio 1-$2 \,\,10^{-5}$, a
$q \bar{q}$ structure with a value $5 \,\,10^{-5}$ \cite{AchIva}
and a $q \bar{q} q \bar{q}$ structure with a value
$2.4 \,\,10^{-4}$ \cite{AchIva}.

The model for $\phi \rightarrow f_0 \gamma$ assumed in Fig.~1
is similar to the one of \cite{AchGabSch} where the production also
proceeds via the kaonic loops. There a $K \bar{K}$ molecule
is assumed for the $f_0$ resonance while here
the realistic $K \bar{K} \rightarrow \pi \pi$ amplitude
of \cite{OllOse1} is used. Emphasis is made in the importance
of going beyond the zero width approximation for the
resonance in \cite{AchGabSch,AchGab}. Our approach automatically
takes this into account since the $K \bar{K} \rightarrow \pi \pi$
amplitude correctly incorporates the width of the $f_0$ resonance
\cite{OllOse1}.

We would also like to warn that the peak of the $f_0$ seen in Fig.~3 cannot
be trivially interpreted as a resonant contribution on top of a background,
since there are important interference effects between the $f_0$ production
and the $\sigma$ background. The strength of the peak comes in our case in
about equal amounts from the real and the imaginary parts of the amplitude for
the process.

The agreement found between our results for the
$\phi\rightarrow \pi^0 \pi^0 \gamma$,
$\phi\rightarrow \pi^0 \eta \gamma$
and experiment provides an important endorsement for the chiral unitary
approach used here. Improvements in the future, reducing the experimental
errors, should put further constraints on avalaible theoretical
approaches for this reaction.

\vspace{3cm}

Acknowledgments:

We would like to acknowledge useful comments from J. A. Oller and from
A. Bramon who called our attention to the recent experimental results
on $\phi\rightarrow \pi^0 \pi^0 \gamma$.
We are grateful to the COE Professorship program of
Monbusho which enabled E.O. to stay at RCNP to perform the present study.
One of us, E.M., wishes to thank the hospitality of the RCNP of the
University of Osaka, and  acknowledges finantial support from the
Ministerio de Educaci\'on y Cultura. This work is partly
supported by DGICYT contract no. PB 96-0753 and by the EEC-TMR
Program Contract No. ERBFMRX-CT98-0169.

\end{document}